\documentclass[%
 reprint,
 superscriptaddress,
 amsmath,amssymb,
 aps,
 prl,
 longbibliography,
lengthcheck,%
]{revtex4-1}

\usepackage{graphicx}
\usepackage{dcolumn}
\usepackage{bm}
\usepackage{color}
\usepackage{ulem}
\usepackage{hyperref}
\usepackage{braket}

\begin{document}

\preprint{APS/123-QED}

\title{
Sign-reversal and non-monotonicity of chirality-related anomalous Hall effect\\in highly conductive metals
}

\author{Ryunosuke Terasawa}
\affiliation{
Department of Physics, Institute of Science Tokyo, Meguro, Tokyo, 152-8551, Japan
}

\author{Masafumi Udagawa}
\affiliation{
Department of Physics, Gakushuin University, Mejiro, Toshima-ku, Tokyo 171-8588, Japan
}

\author{Hiroaki Ishizuka}
\affiliation{
Department of Physics, Institute of Science Tokyo, Meguro, Tokyo, 152-8551, Japan
}

\date{\today}

\begin{abstract}
The non-monotonic temperature dependence and sign reversal of chirality-related anomalous Hall effect in highly conductive metals are studied.
Through the analysis of scattering rate, we find that the non-monotonicity and sign reversal have two major origins: (1) competition between the contribution from short-range and long-range spin correlations and (2) non-monotonic spin correlation in the high-field.
The former mechanism gives rise to non-monotonic temperature dependence in a wide range of electron density and, in some cases, a sign reversal of Hall resistivity as the temperature decreases.
On the other hand, the latter mechanism is responsible for the sign reversal of Hall conductivity in the high field, which sign reversal generally occurs in magnets with antiferromagnetic interactions.
The results demonstrate how the Hall effect reflects local spin correlation and provide insights into the mechanism of non-monotonicity and sign reversal of the anomalous Hall effect by spin chirality.
\end{abstract}

\pacs{
}

\maketitle

{\it Introduction} ---
%
%
Magnetic materials show rich behaviors in their magnetization process, reflecting their magnetic states and excitations, such as the metamagnetic transitions.
Magnetic properties under the external field are often studied through the magnetization curve.
In addition, in magnetic metals, transport phenomena also show rich behaviors reflecting the magnetic properties, some of which are difficult to probe by the magnetization curve.
One such effect is the anomalous Hall effect (AHE) related to 
the scalar spin chirality~\cite{Ye1999a,Ohgushi2000a,Taguchi2001a,Tatara2002a}, which is a distinct mechanism from the ones seen in ferromagnets~\cite{Hall1881a,Karplus1954a,Smit1958a,Berger1970a}. 
Later studies have revealed that the AHE is also sensitive to the vector spin chirality~\cite{Ishizuka2018b,Zhang2018a},
coplanar antiferromagnetic order~\cite{Chen2014a,Lux2020a}, and other non-trivial magnetic states~\cite{Yamaguchi2021a,Terasawa2024a,Mochida2024a}.
In addition to the AHE, other transport phenomena, such as a resistivity minimum~\cite{Udagawa2012a,Wang2016a}, nonreciprocal response~\cite{Yokouchi2017a,Aoki2019a,Ishizuka2020a}, and spin Hall effect~\cite{Ishizuka2013b,Ishizuka2021a,Nakazawa2024a}, are also known to reflect the non-tirival magnetic states.
Such studies have been extended to the local spin correlation of fluctuating spins, in which many experiments study chiral spin fluctuation using the transport phenomena~\cite{Fujishiro2019a,Skoropata2020a,Yang2020a,Fujishiro2021a,Kolincio2021a,Raju2021a,Uchida2021a,Abe2024a,karube2024a}.

\begin{figure}
  \includegraphics[width=\linewidth]{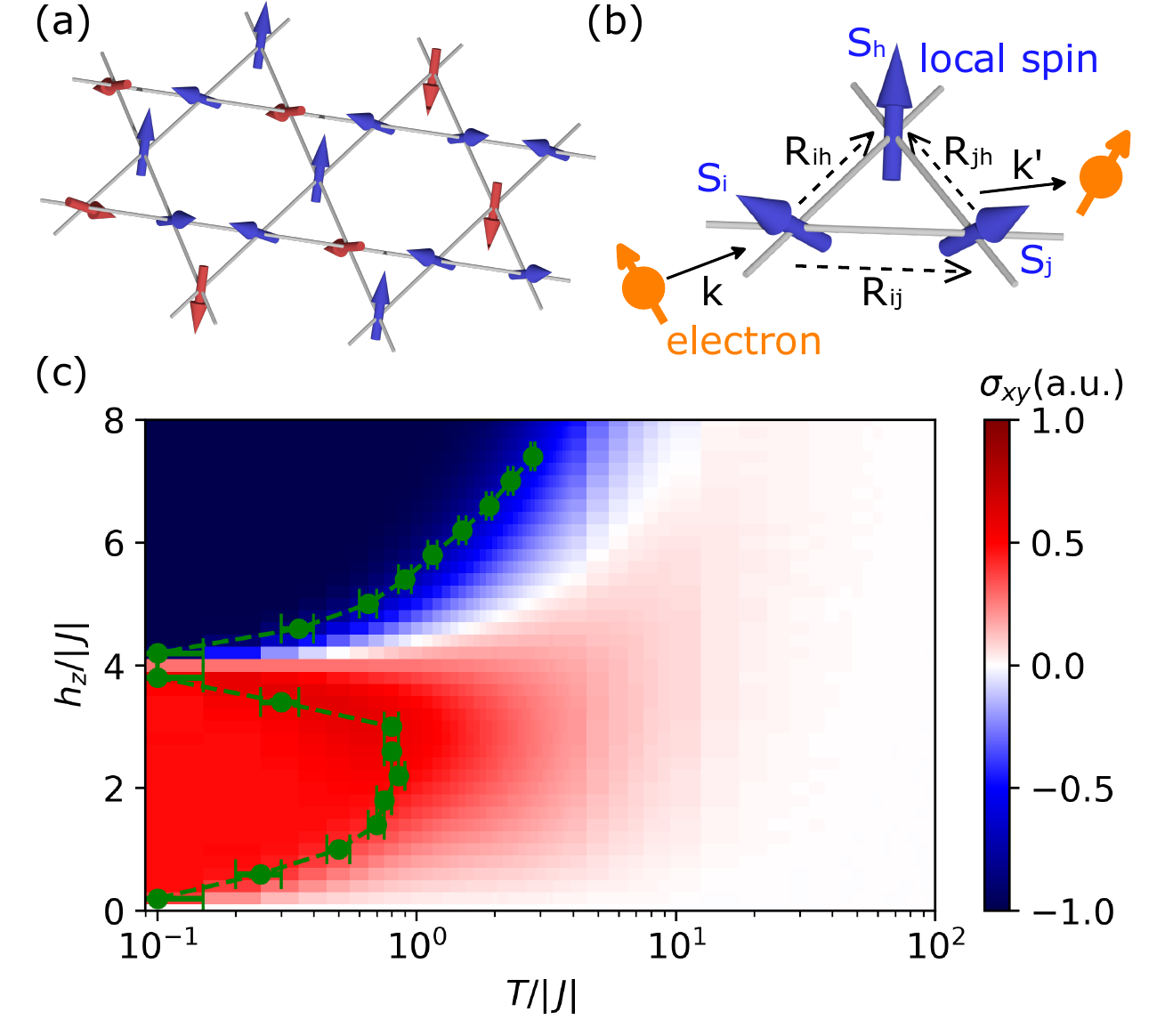}
  \caption{
  Ising spin Kondo lattice model and anomalous Hall effect. (a) A schematic of kagom\'e ice model and (b) skew scattering by multiple spins. (c) Magnetic phase diagram of kagom\'e ice model and the contour plot of anomalous Hall conductivity at $k_Fa=2.5$. The green dots represent the peak of specific heat.
  }\label{fig:model}
\end{figure}

%
%
Among various transport phenomena, the AHE related to spin chirality has become a popular method for detecting chiral magnetic orders, such as skyrmion crystals.
In the AHE, the deviation of anomalous Hall conductivity and magnetization curve is often considered evidence of a chiral magnetic state~\cite{Taguchi2001a,Neubauer2009a}; the deviation is ascribed to the chirality-related AHE.
In experiments, however, the temperature and magnetic field dependence of the chirality-related AHE show rich behaviors.
For instance, a sign reversal of chirality-related AHE at low temperatures is known in moderately clean metals~\cite{Kanazawa2011a,Mayoh2022a}, which is explained by the competition between intrinsic and skew scattering contributions~\cite{Ishizuka2018a}.
Such non-trivial temperature and field dependences were also known in other materials, such as in a highly-conductive metal PdCrO$_2$~\cite{Takatsu2010a,Ok2013a,Jeon2024a}, in which the skew scattering contribution dominates.
However, a systematic understanding of such behaviors in the chirality-related AHE is lacking.

%
%

To explore the origin of non-trivial temperature dependence of anomalous Hall conductivity in highly conductive metals, we studied the AHE in an Ising-spin Kondo-lattice model on a kagom\'e lattice called kagom\'e ice [Fig.~\ref{fig:model}(a)]; it is a basic model for studying chirality-related AHE~\cite{Ohgushi2000a,Ishizuka2013a,Chern2014a,Chern2014a,PhysRevLett.111.036602}and relevant to pyrochlore oxides~\cite{Taguchi2001a,Machida2007a} and heterostructures~\cite{Ohno2024a}.
Using a scattering theory method, we derived a general formula for the anomalous Hall conductivity.
This formula is expressed by the scalar spin chirality and the radial wavefunction of electrons at the Fermi level.
We show that the anomalous Hall conductivity exhibits non-monotonic Fermi wavenumber dependence and sign reversals.
This behavior reflects the nodes of the Bessel function of the first kind.
We also demonstrated that the anomalous Hall conductivity has non-monotonic temperature dependence.
This originates from the competition between short- and long-range spin correlations and/or the non-monotonicity of the spin correlations.
The results indicate that the non-trivial temperature dependence of AHE can occur in highly conductive metals and clarify their physical origins.

{\it Metallic kagom\'e ice} --- 
Here, we consider two-dimensional free electrons coupled to Ising spins on a kagom\'e lattice~\cite{Ishizuka2013a,Chern2014a}. The Hamiltonian reads 
\begin{align}
  H=H_0+H_K,
\end{align}
where, 
\begin{align}
    H_0=\sum_{\bm{k}}\epsilon_{\bm{k}\mu}c_{\mu}^\dagger(\bm{k}) c_{\mu}(\bm{k}),
\end{align}
is the Hamiltonian for free electrons and 
\begin{align}
  H_K=J_K\sum_ic_{\mu}^\dagger(\bm{R}_i)(\bm{S}_{i}\cdot\bm{\sigma})_{\mu\nu}c_{\nu}(\bm{R}_i)
\end{align}
is the Kondo coupling between the classical localized spins and itinerant electrons. Here, $c_\mu(\bm k)$ [$c_\mu^\dagger(\bm k)$] is the annihilation (creation) operator of an electron with momentum $\bm{k}=(k_x,k_y)$ and spin $\mu$, $k=|\bm{k}|$, $\epsilon_{\bm{k}\mu}=k^2/2m$ is the eigenenergy of an electron with momentum $\bm{k}$ and spin $\mu$, $\bm{\sigma}=(\sigma_x,\sigma_y,\sigma_z)$  is the vector of the Pauli matrices, $\bm{S}_{i}=(S_i^x,S_i^y,S_i^z)$ is the $i$th localized moment at position $\bm{R}_i=(R_i^x,R_i^y)$. We assume $\bm{S}_{i}\equiv\tau_i\bm{d}_{i}$ to be an Ising spin, where $\tau_i=\pm1$ and $\bm{d}_{i}$ is the canted easy axis, 
$(-\frac{2\sqrt{2}}{3}, 0, \frac{1}{3})$, $(\frac{\sqrt{2}}{3}, \sqrt{\frac{2}{3}}, \frac{1}{3})$, and $(\frac{\sqrt{2}}{3}, -\sqrt{\frac{2}{3}}, \frac{1}{3})$ 
for the three sublattices of the kagom\'e lattice. In the uniform spin configurations, $\tau_i$ takes $(+,+,+)$ for all-in/all-out, $(+,+,-)$ for FM-I, and $(+,-,-)$ for FM-II, as shown in Fig. 3.
$J_K$ is the strength of exchange coupling between the itinerant electrons and the localized moment.


{\it Boltzmann theory} --- 
To study the AHE arising from the coupling to localized moments, we compute the anomalous Hall conductivity $\sigma_{xy}$ focusing on the skew scattering contribution~\cite{Ishizuka2018a,Ishizuka2021a}.
In the Boltzmann theory, skew scattering is described as an asymmetry of scattering rate.
The scattering rate $W_{\bm{k}\mu\rightarrow \bm{k}'\nu}$ is the rate of electron in $\Ket{\bm k\mu}$ state, the $\mu$th eigenstate of $H_0$ with momentum $\bm k$, being scattered to the state $\Ket{\bm k'\nu}$.
The skew scattering AHE is related to the difference of $W_{\bm{k}\mu\rightarrow \bm{k}'\nu}$ and its inverse process $W_{\bm{k}'\nu\rightarrow \bm{k}\mu}$. 
To study the asymmetry in the scattering rate, we first define the symmetric $w_{\bm{k}\mu\rightarrow \bm{k}'\nu}^+$ and antisymmetric $w_{\bm{k}\mu\rightarrow \bm{k}'\nu}^-$ terms of the scattering rate by 
  $w_{\bm{k}\mu\rightarrow \bm{k}'\nu}^{\pm}=\frac{1}{2}(W_{\bm{k}\mu\rightarrow \bm{k}'\nu}\pm W_{\bm{k}'\nu\rightarrow \bm{k}\mu})$.
In this work, we focus on the asymmetric scattering rate $w_{\bm{k}\mu\rightarrow \bm{k}'\nu}^-$ by the magnetic scattering.

\begin{figure}
  \includegraphics[width=\linewidth]{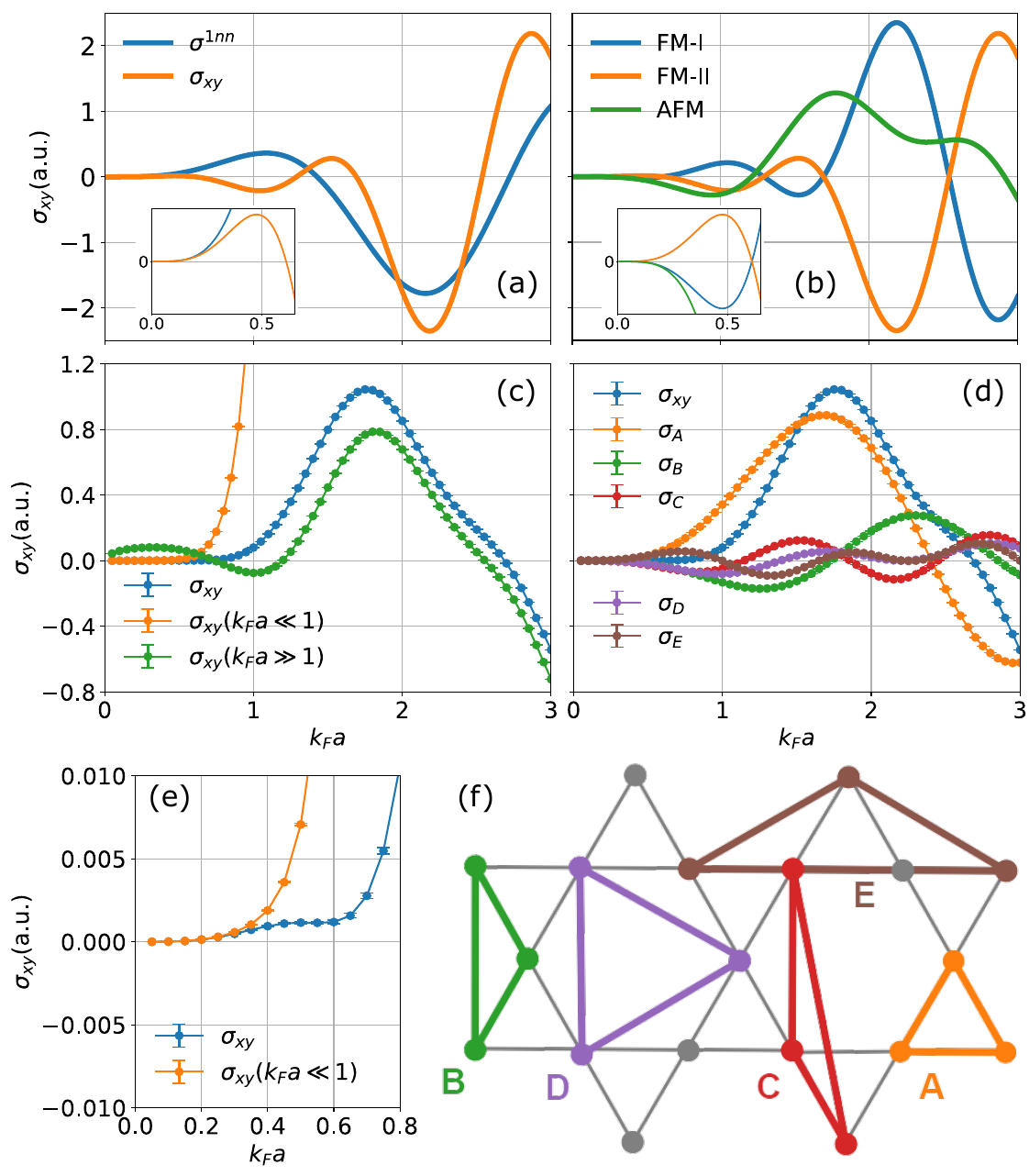}
  \caption{
  Fermi wavenumber $k_F$ dependence of the anomalous Hall conductivity in (a) the All-in/All-out state, (b) FM-I, FM-II, and AFM states, and (c) Chiral kagom\'e ice states; the insets in (a) and (b) show the enlarged view of the small $k_Fa$ region for each figure.
  (d) Contribution of different scattering paths to the Hall conductivity in chiral kagom\'e ice states. (e) An enlarged view of the small $k_Fa$ region of (c). (f) The scattering path involving three spins in kagom\'e lattice.
  }\label{fig:sigmaxy}
\end{figure}

The Hall conductivity is calculated by combining the asymmetric scattering rate (see Supplemental Material~\cite{suppl}) with the semiclassical Boltzmann equation~\cite{Ishizuka2018a,Terasawa2024a}. The Hall conductivity $\sigma_{xy}$ reads
\begin{align}
    \sigma_{xy}=-&\frac{\sigma_{0}k_F^2}{L^2}\sum_{h,i,j}(\bm{S}_{h}\cdot\bm{S}_{i}\times \bm{S}_{j})(\hat{R}_{ih}\times\hat{R}_{jh}\cdot \hat{\bm{z}})\nonumber\\
    &\times J_0(k_FR_{ij})J_1(k_FR_{ih})J_1(k_FR_{jh}).\label{eq:sxy}
\end{align}
Here, $\sigma_0=\frac{\tau^2e^2|m|J_K^3}{2\pi}$, $\bm{R}_{ij}=\bm{R}_i-\bm{R}_j$, $L^2$ is the area of the system, $\hat{\bm{z}}$ is the unit vector along the $z$-axis, $J_n(x)$ $(n=0,1)$ is the Bessel function of the first kind, $k_F$ is Fermi wavenumber, $\tau$ is the relaxation time for symmetric scattering, and $e$ is electric charge.

\begin{figure}
  \includegraphics[width=\linewidth]{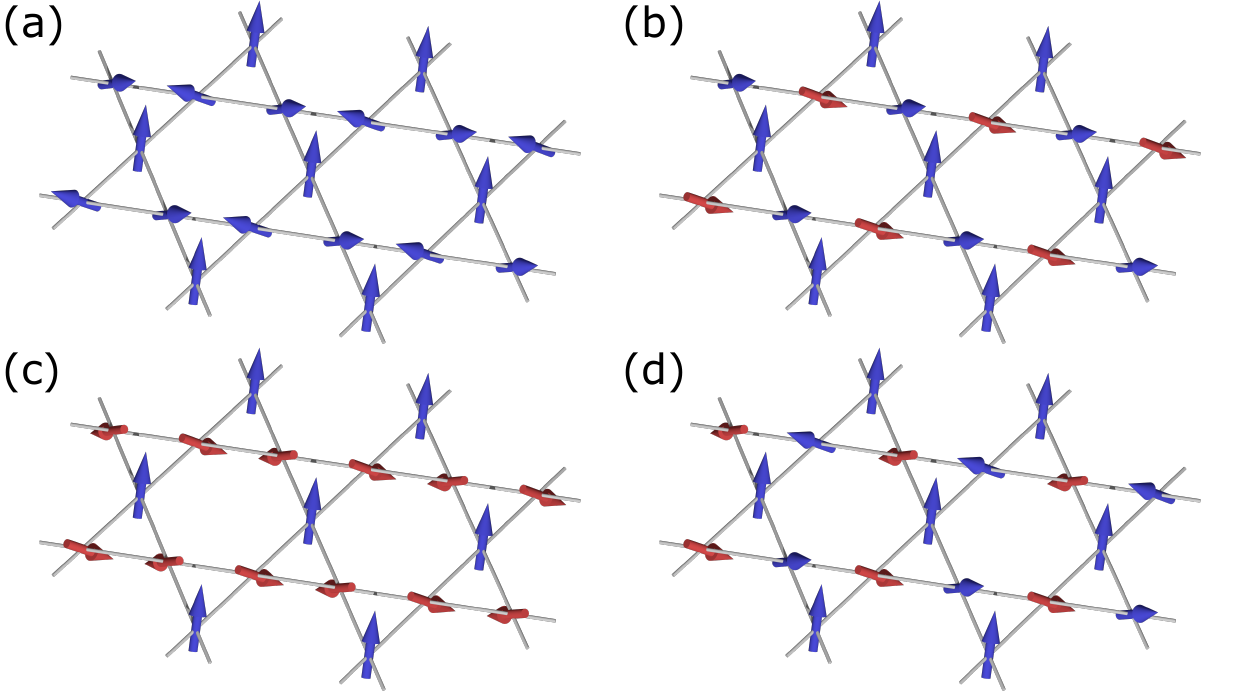}
  \caption{
  Schematics of magnetic orders realized in the kagom\'e ice model: (a) All-in/all-out, (b) FM-I, (c) FM-II, and (d) AFM orders.
  }\label{fig:order}
\end{figure}

{\it Hall conductivity} --- 
Figure~\ref{fig:sigmaxy} shows the Fermi wavenumber $k_F$ dependence of $\sigma_{xy}$ considering up to third-nearest-neighbor contributions; 
Figs.~\ref{fig:sigmaxy}(a)-\ref{fig:sigmaxy}(d) are for all-in/all-out [Fig.~\ref{fig:order}(a)], ordered kagom\'e ice states [see Fig.~\ref{fig:order}(b)-\ref{fig:order}(d)], and the chiral kagom\'e ice state [Fig.~\ref{fig:model}(a)].
The results show non-monotonic $k_F$ dependence and sign reversal, reflecting the nodes in $J_n(x)$.
As evident from Eq.~\eqref{eq:sxy}, the oscillation occurs in the scale of $k_F\sim 1/R_{ij}\sim1/a$ where $a$ is the lattice constant.
Hence, the Hall conductivity changes sensitively with $k_F$ in a metal where $k_Fa$ is comparable to $1$.

For a better understanding of the AHE, let us look into the contributions from different terms in Eq.~\eqref{eq:sxy}.
To this end, we classify the terms in $\sigma_{xy}$ by the scattering process, $\sigma_{xy}=\sigma^{1nn}+\sigma^{2nn}$ where
\begin{align}
    \sigma^{1nn}=&\sigma_A+\sigma_B,\\
    \sigma^{2nn}=&\sigma_C+\sigma_D+\sigma_E,
\end{align}
with $\sigma_a$ ($a=A,\cdots,E$) being the contribution from the scattering path involving three spins in Fig.~\ref{fig:sigmaxy}(f)~\cite{suppl}.
The two terms in $\sigma^{1nn}$ correspond to the contribution from the dominantly nearest-neighbor correlation, and those in $\sigma^{2nn}$ are the contributions involving the second-nearest-neighbor spin correlation; 
the terms in $\sigma^{2nn}$ are limited to $C$, $D$, and $E$ as the scalar spin chirality is zero if two of the three spins belong to the same sublattice due to the common anisotropy axis, $\bm{d}_i$.
In addition, when we consider the third nearest neighbor spins,  the scalar chirality always vanishes because the spins forming a triangle containing the third nearest neighbor spins always contain two or more spins of the same sublattice.
Therefore, there are no contributions involving the third-nearest-neighbor spin correlation.
Figure~\ref{fig:sigmaxy}(a) shows that the trend of $\sigma_{xy}$ is well captured by $\sigma^{1nn}$ for a large $k_Fa\gtrsim1$; similarly, the trend of AHE in the kagom\'e-ice state is well captured by $\sigma_A$ as shown in Fig. 2(d). 
On the other hand, the contribution of $\sigma^{2nn}$ is comparable to that of $\sigma^{1nn}$ when $k_Fa\lesssim1$.
Reflecting the importance of long-range spin correlation, the results for FM-I and AFM ordering in Fig.~\ref{fig:sigmaxy}(b) show different trends
, despite the fact that the contribution from $\sigma_A$ is the same in the two cases.

The relatively large effect of $\sigma^{2nn}$ in the small $k_Fa$ region is understandable from the nature of the Bessel function.
In the $k_Fa\ll1$ limit, the Hall conductivity reads
\begin{align}
    \sigma_{xy}=-\frac{\sigma_{0}k_F^4}{4L^2}\sum_{h,i,j}&(\bm{S}_{h}\cdot\bm{S}_{i}\times \bm{S}_{j})(\bm{R}_{ih}\times \bm{R}_{jh}\cdot \hat{\bm{z}}),
\end{align}
as $J_0(x)\sim1$ and $J_1(x)\sim \frac{x}{2}$ when $x\ll1$.
The equation shows that, in the small $k_Fa$ limit, the contribution from each process is proportional to the area covered by the three spins, $\bm{R}_{ih}\times \bm{R}_{jh}\cdot \hat{\bm{z}}$.
Therefore, the contribution from the further neighbor spins may take over $\sigma_A$ if a long-range spin correlation develops at low temperatures.

The anomalous Hall conductivity in the $k_Fa\lesssim 0.4$ region is well reproduced by further expanding $J_n(x)$ up to ${\cal O}(x^4)$, $J_0(x)\sim1-\frac{x^2}{4}+\frac{x^4}{64}$ and $J_1(x)\sim\frac{x}{2}-\frac{x^3}{16}$,
as shown in Fig.~\ref{fig:sigmaxy}(e).
The narrow window of $k_Fa$ the small $k_Fa$ expansion being valid partly comes from the large $R_{ij}$ in $\sigma^{2nn}$.
Hence, at a high temperature where the spin correlation $\xi$ is limited to $\xi\lesssim a$, the small $k_Fa$ expansion might be valid in a wider range of $k_Fa$.
However, the region of validity of the approximation is limited in a locally correlated phase, such as in the kagom\'e ice state [Fig.~\ref{fig:sigmaxy}(c)].

On the other hand, the relatively small contribution from $\sigma^{2nn}$ in the $k_Fa\gtrsim1$ region is understandable from the asymptotic expansion.
As shown in Fig.~\ref{fig:sigmaxy}(c), the oscillating behavior of Hall conductivity at $k_Fa\gtrsim1$ is well reproduced by the asymptotic expansion of Bessel function about $x\to\infty$.
The asymptotic form of Bessel functions reads $J_0(x)\sim\sqrt{\frac{2}{\pi x}}\cos(x-\frac{\pi}{4})$ and $J_1(x)\sim\sqrt{\frac{2}{\pi x}}\cos(x-\frac{3\pi}{4})$ for $x\gg1$~\cite{Zwillinger2015a}.
Therefore, the contribution from processes involving further-neighbor spins is suppressed by $R^{-3/2}$ where $R$ is the typical distance between the three spins.

{\it Monte Carlo simulation} --- To study how the different contributions are reflected in the temperature dependence,
we calculated the temperature and magnetic field dependence of spin correlation using a Monte Carlo simulation.
To this end, we consider an effective Ising spin model on kagom\'e lattice,
\begin{align}
  H_S=-J\sum_{\langle i,j\rangle}\tau_i\tau_j-h_z\sum_{i}\tau_i,
\end{align}
where $J<0$ is the nearest-neighbor exchange interaction and $h_z$ is the magnetic field.
This model corresponds to the kagom\'e ice model with ferromagnetic nearest-neighbor exchange interaction of $-3J$ between canted Ising spins, and the magnetic field $3h_z$ perpendicular to the plane.

\begin{figure}
  \includegraphics[width=\linewidth]{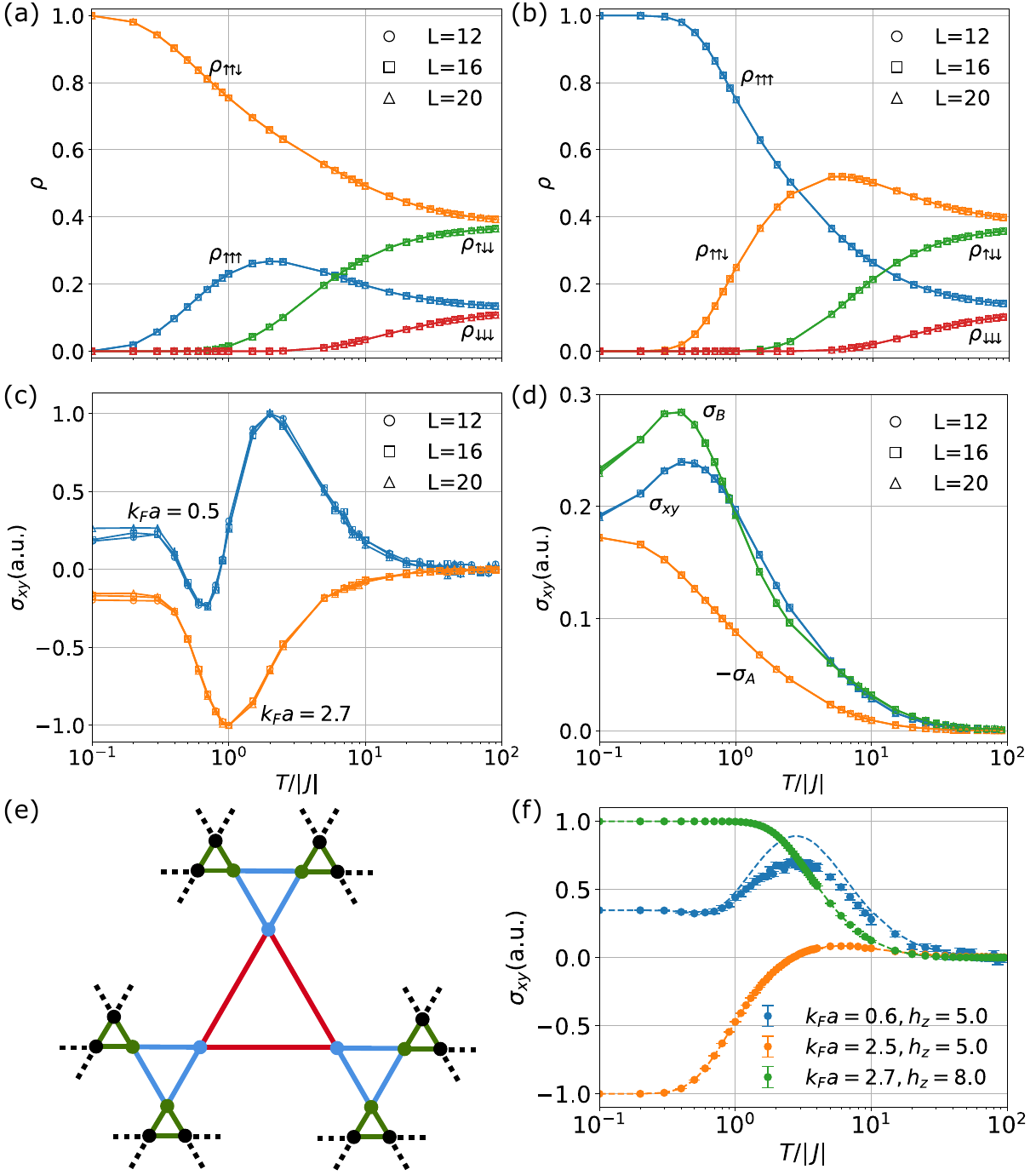}
  \caption{
  Monte Carlo simulation of the kagom\'e ice model under the perpendicular magnetic field.
  The temperature dependence of the population of triangles with all-up ($\rho_{\uparrow\uparrow\uparrow}$), two-up-one-down ($\rho_{\uparrow\uparrow\downarrow}$), one-up-two-down ($\rho_{\uparrow\downarrow\downarrow}$), all-down ($\rho_{\downarrow\downarrow\downarrow}$) (a) at $h_z=3.5$ and (b) at $h_z=5.0$.
  The temperature dependence of anomalous Hall conductivity (c) at $h_z=2.0,k_Fa=0.5, 2.7$ and (d) at $h_z=3.4,k_Fa=2.5$.
  (e) A schematic of the Husimi tree.
  (f) The dots are the results of Monte Carlo simulations, and the dashed lines are the results of the Husimi tree.
  The simulation was performed with a system of $N=3\times20^2$ sites, using $10^8$ Monte Carlo steps each for equilibration and measurement.
  }\label{fig:MC}
\end{figure}

Under the ambient magnetic field $|h_z/J|<4$, the low-temperature state of this model is a locally correlated disorder state~\cite{Udagawa2002a,PhysRevB.68.064411,Udagawa2021}.
At $h_z=0$, any spin configurations with all triangles in the two-up-one-down or one-up-two-down state are degenerate; it is a classical spin liquid state.
Under a weak magnetic field $0<h_z<4|J|$, on the other hand, the magnetic field partially lifts the macroscopic degeneracy at $h_z=0$ as shown in Fig.~\ref{fig:MC}(a).
However, the spin configurations with all triangles in the two-up-one-down state are degenerate; this is the chiral kagom\'e ice (CKI) state [Fig.~\ref{fig:model}(a)].
The ground state is the all-in/all-out (AIAO) state [Fig.~\ref{fig:order}(a)] for $h_z>4|J|$ as in Fig.~\ref{fig:MC}(b). 

The temperature and magnetic field dependence of AHE is shown in Fig.~\ref{fig:model}(c).
For $k_Fa=2.5$, the sign of the Hall conductivity is positive in the CKI state and negative in the AIAO state at low temperatures.
It is consistent with the strong coupling theory~\cite{Ishizuka2013a} and that observed in an experiment~\cite{Ohno2024a}.
However, in general, the magnetic field and the temperature dependence show different behaviors depending on $k_F$, which cannot be explained by the temperature dependence of spin chirality for the smallest triangles $A$.
Moreover, the temperature dependence of $\sigma_{xy}$ is often non-monotonic with the temperature as shown in Fig.~\ref{fig:MC}(c).
In the last, we note that the sign of $\sigma_{xy}$ for $k_Fa=2.5$ reflects the scalar spin chirality of triangle $B$, not $A$, as shown in Fig.~\ref{fig:MC}(d).
The result illustrates the importance of considering the contributions from further-neighbor spin correlation.

\begin{figure}
  \includegraphics[width=\linewidth]{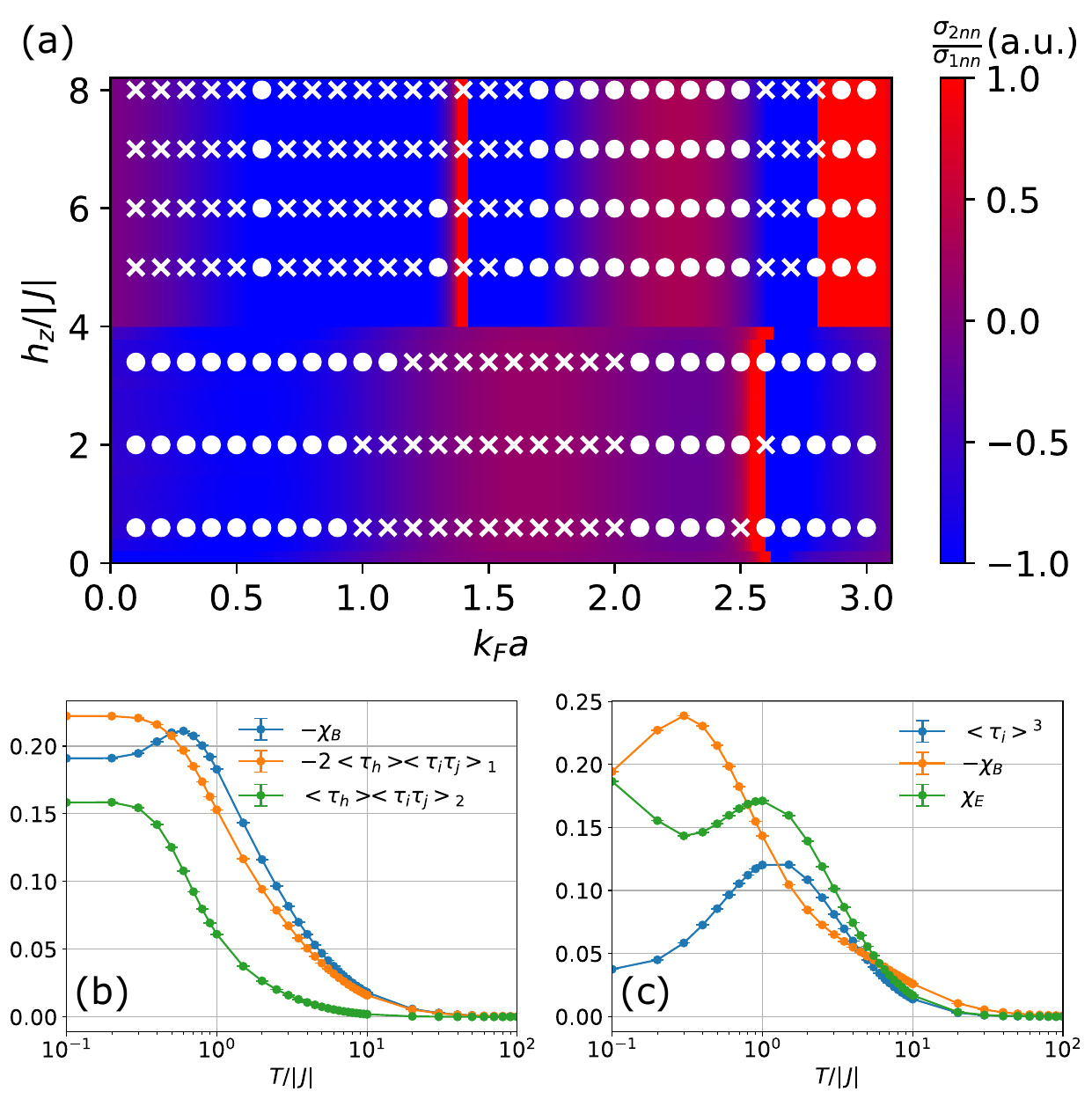}
  \caption{
  (a) The $k_F$ and $h_z$ dependence of $\sigma_{2nn}/\sigma_{1nn}$ at $T=0$. The $\bigcirc$($\times$) denotes that $\sigma_{xy}$ behaves non-monotonically (monotonically) as the temperature changes at the parameter. $T$ dependence of (b) $\chi_B, \langle \tau_h \rangle \langle \tau_i \tau_j \rangle_1, \langle \tau_h \rangle \langle \tau_i \tau_j \rangle_2$ at $h_z=1.5$ and (c) $\chi_B, \chi_E, \langle \tau_i \rangle^3$ at $h_z=3.5$.
  The simulation was performed with a system of $N=3\times20^2$ sites, using $10^8$ Monte Carlo steps each for equilibration and measurement.
  }\label{fig:summary}
\end{figure}

{\it Non-monotonicity at low magnetic fields} ---
%
%
To understand the origin of the non-monotonic temperature dependence in Figs.~\ref{fig:MC}(c) and \ref{fig:MC}(d), we plotted $\sigma^{2nn}/\sigma^{1nn}$ with respect to the $k_F$ and magnetic field in Fig.~\ref{fig:summary}(a).
The non-monotonic temperature dependence at low magnetic fields, including sign reversal of $\sigma_{xy}$ in some cases (see Supplemental Material~\cite{suppl}), occurs in $k_Fa\lesssim1$ and $k_Fa\gtrsim2$, the $k_F$ at which $\sigma^{2nn}/\sigma^{1nn}$ is mostly negative.
The agreement implies that the non-monotonicity is a consequence of the competition between $\sigma^{1nn}$ and $\sigma^{2nn}$.
The spin chirality of the triangles that contribute to $\sigma_{xy}$ shows different temperature dependences; often $\sigma^{1nn}\gg\sigma^{2nn}$ at high temperatures and large $k_Fa$.
In such a case, the sign of AHE at high temperatures reflects the sign of $\sigma^{1nn}$.
However, once $\sigma^{2nn}$ starts to increase at low temperatures, the increase induces non-monotonic temperature dependence.
Consequently, the non-monotonic temperature dependence appears when the contribution from the further-neighbor spin correlation is large and $\sigma^{2nn}/\sigma^{1nn}<0$.

The non-monotonic temperature dependence of $\chi_B$ in $0\lesssim h_z \lesssim2|J|$ is also attributed to a similar competition between short- and long-range two-spin correlations in the two scattering paths in $B$ [Fig.~\ref{fig:summary}(b)].
We note that the spin correlation of the spin-ice-like states is well described by an effective Gaussian theory~\cite{Isakov2004}.
In the Gaussian theory, the three-spin correlation is given as the sum of $\langle \tau_i\rangle\langle \tau_j\tau_k\rangle$, where $i$, $j$, and $k$ are the indices of three spins on a triangle; at high temperatures, the same conclusion derives from the high-temperature expansion.
For triangles like $B$, in which the distance between spins differs depending on spin pairs, long-range correlations decay faster than short-range ones, as shown in Fig.~\ref{fig:summary}(b).
Using the Husimi tree, we find that the nearest-neighbor spin correlation is $\langle \tau_i \tau_j \rangle_{1} = \beta J$ and second-nearest-neighbor spin correlation $\langle \tau_i \tau_j \rangle_{2} = \beta^2 (J^2 + h^2)$~\cite{suppl}.
As $\langle \tau_i \tau_j \rangle_1<0$ and $\langle \tau_i \tau_j \rangle_2>0$ when $J < 0$, $\chi_B$ becomes non-monotonic at low temperatures in which long-range correlation develops.
The non-monotonicity of $\chi_B$ is reflected in $\sigma_{xy}$ near $k_Fa=2.5$ where the influence of $\chi_B$ becomes relatively large compared to $\sigma_A$, as shown in Fig.~\ref{fig:MC}(d).

Finally, we note that the non-monotonic temperature dependence of magnetization $\langle \tau_i\rangle$ contributes to the non-monotonicity of $\sigma_{xy}$ near $h=4|J|$.
As shown in Fig.~\ref{fig:summary}(c), a non-monotonic temperature dependence of $\langle \tau_i\rangle$ is often seen in the range $2|J|\lesssim h_z \lesssim4|J|$.
It is likely related to the fact that the lowest excited state is the all-up state for $2|J|\lesssim h_z \lesssim4|J|$.
Calculating the magnetization at low temperatures using the Husimi tree [Fig.~\ref{fig:MC}(e)], which is known to be exactly solvable ~\cite{Husimi1950a,Harary1953a}, we find
\begin{align}
\langle \tau_i \rangle &= \frac{1}{3} + \frac{2\sqrt{2}}{9}u - \frac{\sqrt{2}}{3}d - \frac{2}{9}l,
\end{align}
where $u=e^{\beta(4J+h_z)}, d=e^{-\beta h_z}, l=e^{\beta(4J-2h_z)}$~\cite{suppl}.
This equation indicates that $\langle \tau_i\rangle$ becomes non-monotonic at low temperatures when $u > d$, i.e., in the range $2|J|\lesssim h_z \lesssim4|J|$.
Such non-monotonicity of magnetization also takes part in the non-monotonic temperature dependence near $h=4|J|$.

{\it Non-monotonicity at high magnetic fields} --- 
%
On the other hand, the sign reversal of $\sigma_{xy}$ at high fields $h>4|J|$ [see Fig.~\ref{fig:model}(c)] cannot be explained by the competition of $\sigma^{1nn}$ and $\sigma^{2nn}$ as inferred from Fig.~\ref{fig:summary}(a).
For a better understanding of the origin of non-monotonicity at high magnetic fields [Fig.~\ref{fig:MC}(f)], we calculate the three-spin correlation with high-temperature expansion.
The thermal average reads
\begin{align}
  \langle \tau_i\rangle=&\beta h_z(1+4\beta J),\\
  \langle \tau_i\tau_j\rangle=&(\beta h_z)^2+\beta J,\\
  \langle \tau_i\tau_j\tau_k\rangle=&(\beta h_z)^3+3\beta^2h_zJ,\label{eq:hiT-chirality}
\end{align}
where $\langle O\rangle$ represents the thermal average of $O$~\cite{suppl}.
Equation~\eqref{eq:hiT-chirality} shows that the sign of chirality changes at $T=-h_z^2/3J$ when $J<0$.
As implied from the derivation, which uses the fact that the spins are of Ising type and the exchange interaction is antiferromagnetic, the sign reversal of $\langle \tau_i\tau_j\tau_k\rangle$ is a general feature of antiferromagnetic Ising models.


{\it Summary} --- 
In this paper, we studied the temperature dependence of AHE induced by a skew scattering mechanism in a two-dimensional free electron coupled to Ising spins on a kagom\'e lattice.
Using a scattering theory method, we derived a general formula for the anomalous Hall conductivity expressed by a combination of the scalar spin chirality and the Bessel functions of the first kind.
We found that the anomalous Hall conductivity exhibits non-monotonic Fermi wavenumber dependence and sign reversal, reflecting the radial wavelength of electric wavefunctions at the Fermi level.
Depending on the value of the Fermi wavenumber and magnetic field, the temperature dependence of Hall conductivity varies [Fig.~\ref{fig:summary}(a)].
At low magnetic fields ($0\lesssim h_z \lesssim4|J|$), non-monotonic behavior caused by the competition between $\sigma^{1nn}$ and $\sigma^{2nn}$ and/or the non-monotonicity of spin correlation is observed around $k_Fa\lesssim1$ and $k_Fa\gtrsim2$.
At high magnetic fields ($h_z\gtrsim4|J|$), a sign reversal of $\sigma_{xy}$ is observed at $k_Fa\gtrsim1.7$, reflecting the sign change of chirality at high temperatures.
These results demonstrate the non-trivial temperature dependence of chirality-related AHE by skew scattering, which should dominate the AHE in highly-conductive metals.

\acknowledgements
We are grateful to J. Fujii for providing Python scripts for analyzing the Monte Carlo results.
This work is supported by JSPS KAKENHI (Grant Numbers JP19K14649, JP23K03275, JP20H05655, JP22H01147, JP23K22418) and JST PRESTO (Grant No. JPMJPR2452).

%
%
%
%
%

\end{document}